\documentclass[%
 reprint,
 amsmath,amssymb,
 aps,
floatfix
]{revtex4-2}

\usepackage[dvipsnames]{xcolor}

\usepackage{graphicx}
\usepackage{dcolumn}
\usepackage{bm}

\usepackage{todonotes}
\renewcommand{\d}{\,\text{d}}

\definecolor{myellow}{rgb}{1., 1., 0.6}

\makeatletter
\newcommand{\fmarki}{*}
\newcommand{\fmarkii}{\ensuremath{\dagger}}
                
\def\@fnsymbol#1{{\ifcase#1\or \fmarki\or \fmarkii \else\@ctrerr\fi}}
\makeatother

\renewcommand{\fmarki}{}
\renewcommand{\fmarkii}{}

\usepackage[normalem]{ulem} 
\usepackage{bbm}

\usepackage{physics} 
\usepackage[T1]{fontenc} 

\begin{document}

\title{Atomic physics on a 50 nm scale: Realization of a bilayer system of dipolar atoms}

\author{Li Du$^{*,\dag}$}
\thanks{$*$\ \ These authors contributed equally.}
\author{Pierre Barral$^{*}$}
\thanks{$*$\ \ These authors contributed equally.}
\author{Michael Cantara$^{*}$}
\thanks{$*$\ \ These authors contributed equally.}
\author{Julius de Hond}
\author{Yu-Kun Lu}
\author{Wolfgang Ketterle}
\affiliation{
Research Laboratory of Electronics, MIT-Harvard Center for Ultracold Atoms, and Department of Physics, Massachusetts Institute of Technology, Cambridge, Massachusetts 02139, USA
}
\email{$\dag$\ \ E-mail: lidu@mit.edu}

\date{\today}

\clearpage
\maketitle

\textbf{Atomic physics has greatly advanced quantum science, mainly due to the ability to control the position and internal quantum state of atoms with high precision, often at the quantum limit.  The dominant tool for this is laser light, which can structure and localize atoms in space (e.g., in optical tweezers, optical lattices, 1D tubes or 2D planes).  Due to the diffraction limit of light, the natural length scale for most experiments with atoms is on the order of 500 nm or larger.  Here we implement a new super-resolution technique which localizes and arranges atoms on a sub-50 nm scale, without any fundamental limit in resolution. We demonstrate this technique by creating a bilayer of dysprosium atoms, mapping out the atomic density distribution with sub-10 nm resolution, and observing dipolar interactions between two physically separated layers via interlayer sympathetic cooling and coupled collective excitations.  At 50 nm,  dipolar interactions are 1,000 times stronger than at 500 nm. For two atoms in optical tweezers,  this should enable purely magnetic dipolar gates with kHz speed.}\\

A major frontier in many-body physics is the realization and study of strongly-correlated quantum phases \cite{keimer2017physics, andrei2020graphene, lewenstein2012ultracold}. In ultracold atomic systems, the typical short-range contact interaction has led to the creation of a variety of exotic quantum phases \cite{bloch2008many, lewenstein2012ultracold}. However, a wide range of quantum phenomena require long-range dipolar interactions
\cite{chomaz2022dipolar, lahaye2009physics, baranov2012condensed}.
But even for the most magnetic atoms such as erbium (Er) and dysprosium (Dy), the magnetic dipole-dipole interaction is rather weak. For Dy, with a magnetic dipole moment of 10 Bohr magneton ($\mu_B$), the dipolar interaction at 500 nm distance is only $h \times 20$~Hz, $h$ being the Planck's constant. 
Although several new forms of matter could be realized with magnetic atoms \cite{chomaz2022dipolar}, there are major efforts to harness the much stronger interactions of polar molecules \cite{gadway2016strongly, moses2017new} and Rydberg atoms \cite{browaeys2020many}.  The electric dipolar interaction of molecules (at 3 Debye) can be 1,000 times stronger than magnetic dipolar interaction (at 10 $\mu_B$).  Here we show how this factor of 1,000 can be compensated for by decreasing the distance between two magnetic atoms to 50~nm. To pursue dipolar physics with atoms has the major advantage that it is simpler to cool atoms, and that atoms have more favorable collisional properties --- so far, polar molecules have yet to be Bose condensed.

\begin{figure*}[htbp]
\includegraphics[width=1.14\textwidth]{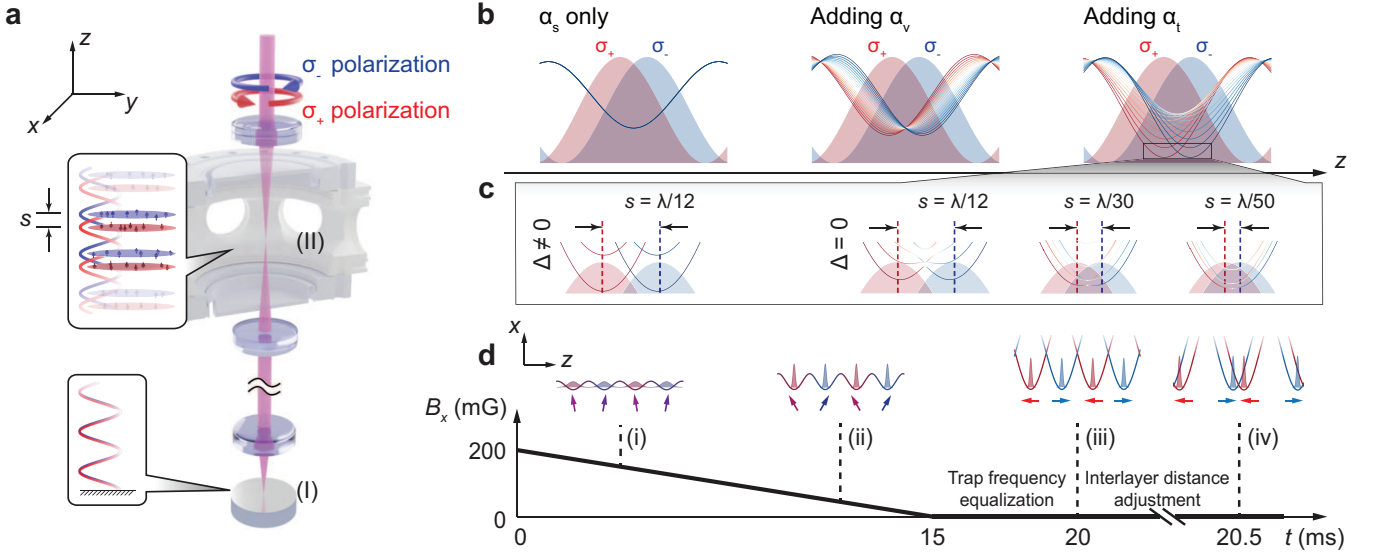}\\
    \caption{\textbf{Creation of the subwavelength bilayer array.}
    \textbf{a}, Experimental setup. Two overlapping laser beams with opposite circular polarizations $\sigma_+$ and $\sigma_-$ are retro-reflected by mirror (I) to form two optical standing waves. The two standing waves are displaced at the position of the atoms (II), controlled by the frequency offset $\Delta$ between the two laser beams. Dy atoms in this configuration form an array of pancake-shaped bilayers of head-to-head dipoles with adjustable interlayer distance $s$. 
    \textbf{b}, Contributions of different polarizability components.  Solid lines denote adiabatic potential curves for different $m_J$ states ($-8 \leq m_J \leq 8$ --- blue to red, assuming red detuning), and the shaded regions refer to the intensities of the $\sigma_\pm$ light for a particular interlayer separation.  
    \textbf{c}, If the two laser beams have the same frequency $\Delta = 0$, the off-diagonal part of the tensor polarizability mixes spin states. As a result, the two minima merge into a single minimum for small separation $s$. This is avoided in our experiment by using two different frequencies for the $\sigma_\pm$ light. The color of the curves indicate the $m_J$ character of the adiabatic eigenstates.
    \textbf{d}, Adiabatic loading of the bilayer array. (i) Starting with the optical potential in the interlaced configuration in the presence of a dominating transverse magnetic field $B_x = 200~\rm{mG}$, the atoms are initialized in the $m_J = -8$ spin state along the $\mathbf{x}$ direction. (ii) As $B_x$ is ramped down in $15~\rm{ms}$, the light shift dominates over the Zeeman shift, thereby adiabatically loading the bilayer array.
    (iii) The power of the $\sigma_+$ and $\sigma_-$ potentials are adjusted for identical trap frequencies. (iv) The interlayer distance is adjusted to designated values in $0.5~\rm{ms}$.
    }
\label{fig:potential-cartoon}
\end{figure*}

It has been a long-standing goal to create potentials with subwavelength components to enhance tunneling and interaction strengths. Many schemes have been suggested \cite{caldwell2020enhancing, caldwell2021general, lkacki2016nanoscale, nascimbene2015dynamic, kruckenhauser2020quantum} and methods such as dark states \cite{Wang18, mcdonald2019superresolution}, RF-photon dressing \cite{lundblad2008atoms}, stroboscopic techniques \cite{Tsui20}, and Fourier synthesis with multiphoton processes \cite{ritt2006fourier, anderson2020realization} have been demonstrated. So far, none of these methods have been widely used due to additional heating and limited coherence time, and often they reduced the atomic spacing by a factor of only two or three. Our super-resolution method has no fundamental limit.  It is based on the key concept of super-resolution microscopy that one can determine the center of a diffraction-limited Airy disk much more precisely than the diffraction limit itself. Similarly, a deep optical lattice or a strong tweezer beam can localize an atom to 10 nm \cite{forster2009microwave, kaufman2012cooling}, limited only by available power and heating from spontaneous light scattering.  In super-resolution microscopy, molecules are imaged sequentially, whereas for trapping atoms, simultaneous confinement on a sub-wavelength scale is required. One possible solution is to trap two different kinds of atoms with two different colors of light. But usually for quantum science, one needs identical atoms. The strategy implemented here uses two opposite spin states of Dy, and two different polarizations of light at different frequencies --- a dual polarization and dual frequency super-resolution scheme. Unlike spin-1/2 and alkali atoms, ground-state Dy has a strong tensor polarizability \footnote{Spin-1/2 atoms and alkalis for detunings larger than the excited-state hyperfine splitting share the property that they have only a scalar and vector polarizability. In general, atoms have a vector or tensor polarizability when light scattering can change their angular momentum by $\hbar$ or $2 \hbar$, respectively \cite{LeKien13, cui2013synthetic}.} which has a good and a bad side. It causes detrimental two-photon Raman couplings between $m_J$ states, which are suppressed by the frequency offset between the two optical potentials. With that, the remaining diagonal part of the tensor couplings makes our scheme much more robust since it creates, for $^{162}\rm{Dy}$, an isolated two-state Hilbert space for $m_J = \pm 8$ spin states with a big energy gap to all the other 15 spin states. This is the main new concept of our scheme.

Spin-dependent potentials have been realized with rubidium (Rb) \cite{Mandel2003, Yang17, soltan2011multi, deHond22, gadway2010superfluidity} and cesium (Cs) \cite{forster2009microwave, belmechri2013microwave}. However, strong spin-dependent potentials for alkali atoms require near-resonant light with detunings smaller than the fine-structure splitting $\Delta_{\rm{FS}}$, causing severe heating due to spontaneous emission. The ratio of the spontaneous emission rate $\Gamma$ to potential depth is limited to $\Gamma/\Delta_{\rm{FS}}$ \cite{mckay2010thermometry}, which is on the order of $10^{-6}$ for Rb and Cs. More specifically, for Rb, the trap depth used here (1000 $E_{\rm{recoil}}$) would cause a spontaneous emission time of only 15 ms \cite{mckay2010thermometry}. In contrast, very deep spin-dependent potentials can be realized with Dy with negligible spontaneous emission thanks to Dy's spin-orbit coupling in the ground state. Furthermore, with a magnetic dipole moment of only $1~\mu_B$, the dipolar interaction for alkali atoms is 100 times weaker than for Dy. Therefore, previous work on alkalis has used spin-dependent forces to control the overlap between sites with spin up and down \cite{forster2009microwave, deHond22}, but not to study interactions between non-overlapping sites.

A systematic illustration of the super-resolution scheme is shown in Fig.~\ref{fig:potential-cartoon}(b), demonstrating a bilayer potential created by two optical standing waves of $\sigma_+$ and $\sigma_-$ polarizations with a small spatial displacement $s$. This illustration also applies to the case of spin-dependent optical tweezers. The figure shows the adiabatic potentials of all 17 spin states (taking quantization axis along $\mathbf{z}$ direction in the lab frame), with different polarizability components taken into account. With only a scalar polarizability $\alpha_s$, the AC Stark shifts are the same for all 17 $m_J$ states, so there is only one potential minimum. The vector polarizability $\alpha_v$ leads to AC Stark shifts that are linear in $m_J$, and therefore can be regarded as a Zeeman shift caused by a fictitious sinusoidal magnetic field --- it lifts the degeneracy except for points  where the fictitious magnetic fields from the $\sigma_+$ and the $\sigma_-$ standing waves cancel. This creates a double well potential even for arbitrarily small displacement of the standing waves. However, small transverse magnetic fields would couple the degenerate states, leading to mixing among many $m_J$ states and causing losses by dipolar relaxation. This is where the tensor polarizability $\alpha_t$ makes a qualitative difference.  The diagonal part of the tensor light-atom interaction (which has an $m_J^2$ dependence) partially lifts the degeneracy, and results in potentials where only states with the same $|m_J|$ cross in the middle. Hence the $m_J = \pm 8$ ground states  are separated from all other states by a large gap and are coupled by transverse fields only in sixteenth order. Note that the  $m_J = \pm 8$ states are the local ground states of the $\sigma_\pm$ potential minima, and therefore inelastic two-body losses are prevented in each of the layers.

Although the tensor polarizability $\alpha_t$ provides robustness against transverse magnetic fields, it allows for two-photon Raman processes with $\Delta m_J= \pm2$ using one $\sigma_+$ and one $\sigma_-$ photon.  Fig.~\ref{fig:potential-cartoon}(c) shows the effect of the resonant Raman process due to off-diagonal tensor couplings when both polarization components have the same frequency. This is the situation when the $\sigma_+$ and the $\sigma_-$ standing waves are created by retro-reflecting a single beam in the lin-$\vartheta$-lin geometry, as often used for alkalis (e.g.~\cite{Mandel2003,Yang17,deHond22}). For Dy, the Raman couplings induce tunneling and they weaken the potential minima when their separation becomes smaller than $\lambda /10$, where $\lambda$ is the wavelength of the light. For displacements of the standing waves of less than $\lambda /30$, the double minima has merged to a single minimum. We eliminate the Raman coupling by offsetting the frequencies for the $\sigma_+$ and  $\sigma_-$ optical standing waves by more than 300~MHz, much larger than the AC Stark shifts, which makes the two-photon Raman process off-resonant \footnote{If the Raman couplings were eliminated using an external magnetic field, it breaks the symmetry of the $m_J =\pm 8$ states and causes rapid dipolar relaxation in the  $m_J = +8$ layer.}. The conclusion is that the dual polarization and dual frequency scheme isolates the Hilbert space for the $m_J = \pm 8$ spin states and creates a double minimum potential which is not flattened out even for very small separations of the two minima.

Dy, with its high angular momentum of $J=8$ in the ground state, is the ideal atom for this scheme. For a $J=8 \rightarrow J'=9$ transition, the $m_J =8$ state has a transition strength ratio of 153 between the $\sigma_+$ and $\sigma_-$ transitions \footnote{Depending on detuning, this contrast doesn't fully translate into the AC Stark shift because of the scalar background polarizability due to the strong transition triplet at 421, 419 and 405 nm.}.  For atoms with $J=1$ (2), the ratio is only 6 (15). Therefore, this stretched transition in Dy is very similar to a hypothetically isolated $J=1/2 \rightarrow J' = 1/2$ transition, where the spin-up state sees only $\sigma_-$ light and vice versa. Dy is even more ideal than the $J=1/2$ case where spin-up and down states are directly connected by possible one- or two-body couplings (e.g.\ transverse magnetic fields, dipolar relaxation), whereas those couplings act only in sixteenth or eighth order in our Dy scheme.  The robustness of the scheme comes from the AC Stark shifts due to the tensor polarizability.

Experimentally, a stack of bilayers is created by superimposing red-detuned optical standing waves with $\sigma_+$ and $\sigma_-$ polarizations operating near the Dy narrowline transition at $741~\rm{nm}$ (linewidth $\Gamma/2\pi = 1.78~\rm{kHz}$) \cite{Lu11}. The two optical beams are delivered through the same polarization-maintaining fiber, such that they share the same transverse Gaussian mode. 
The frequency of the $\sigma_-$ standing wave can be dynamically tuned using a double-passed phased array acousto-optic deflector, leading to a precise control of the interlayer distance $s$ with a sensitivity of $4.7~\rm{nm/MHz}$ (see Methods).

\begin{figure}[bp]
    \centering
    \includegraphics[width=\columnwidth]{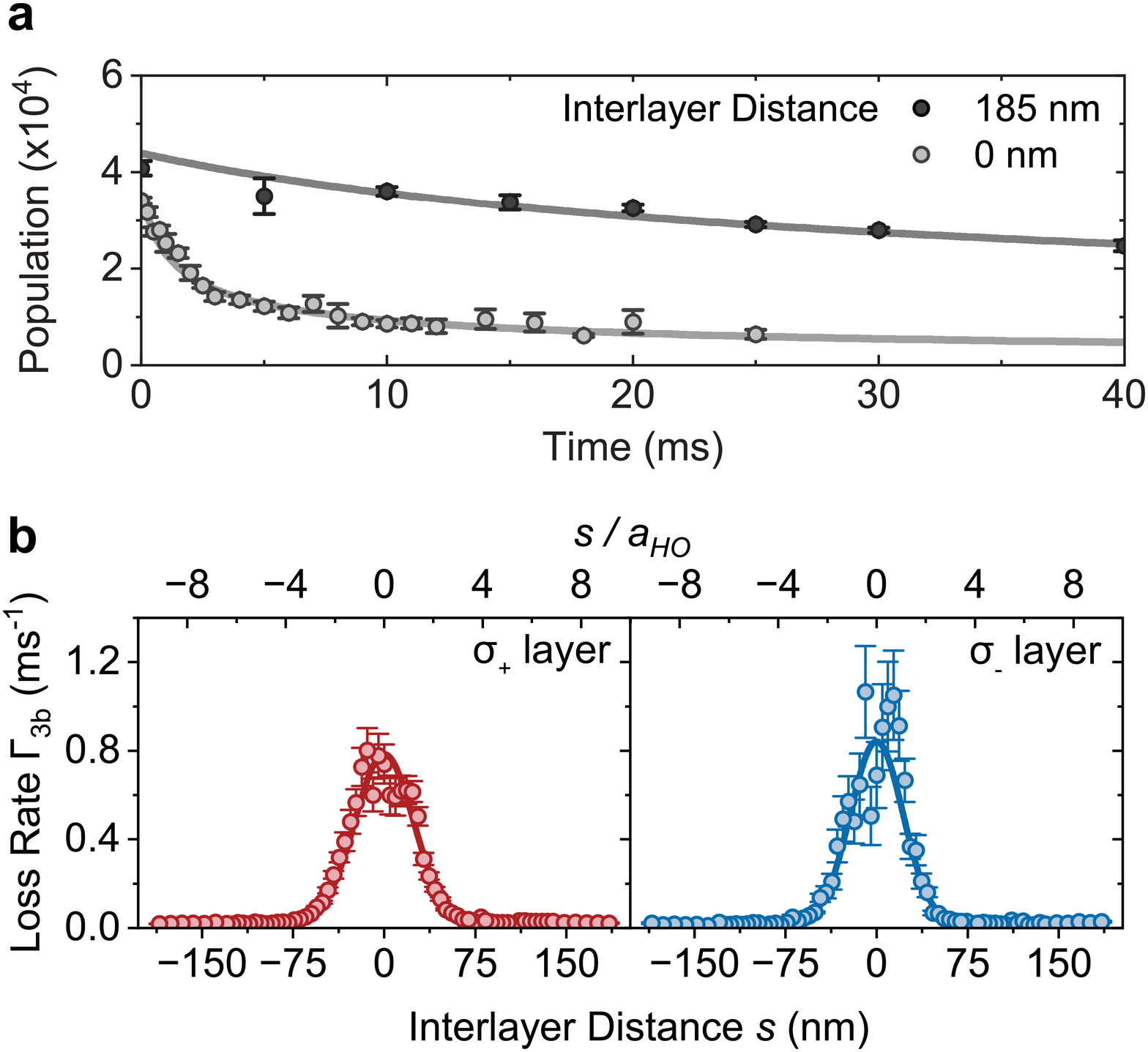}
    \caption{\textbf{Demonstration of controlling the interlayer distance on a 10 nm scale by recording atom loss as a function of layer separation.}
    \textbf{a}, Evolution of the population in $\sigma_+$ layers at two different interlayer distances $s=185~\rm{nm}$ and $0~\rm{nm}$. The loss is much faster when the layers are overlapped. Initial loss rates $\Gamma_{\rm{3b}}$ are obtained from the fits to the decay curves.
    \textbf{b}, Gaussian fits of the initial loss rates $\Gamma_{\rm{3b}}$ to the interlayer distances $s$ according to Eq.~(\ref{eq:LossRate}) (solid lines) provide a value of  $\sigma_z = 18.6~\rm{nm}$ for the layer thickness.
    }
    \label{fig:loss-measurement}
\end{figure}

The ground state of the bilayer is loaded using an adiabatic transfer method, as depicted in Fig.~\ref{fig:potential-cartoon}(d). First,  $m_J = -8$ atoms are prepared in a magnetic field with a transverse component $B_x = 200~\rm{mG}$ and an axial component $B_z$ around $10~\rm{mG}$. We then ramp up the $\sigma_+$ and $\sigma_-$ standing waves in the interlaced configuration (i.e. with $s = \lambda/4$) in $100~\rm{ms}$, loading all layers with atoms aligned with the $x$ axis (Fig.~\ref{fig:potential-cartoon}(d, i)).  By ramping down the transverse magnetic component $B_x$ in $15~\rm{ms}$, the potential depth increases while a bilayer array is formed with dipoles that are aligned head-to-head (Fig.~\ref{fig:potential-cartoon}(d, ii-iii)). We ensure balanced loading by making sure that the energy offset between the minima of the  $\sigma_+$ and $\sigma_-$ potentials $\delta = U_+ - U_- +E_Z=0$, where $U_\pm$ are the AC Starks shift of the layers, and the differential Zeeman energy is given by $E_Z=16g_J \mu_B B_z$. It is crucial that the atoms stay in their local ground state throughout the experiment to prevent losses and heating due to dipolar relaxation. Therefore, the Zeeman shifts caused by the external magnetic field $B_z$ have to be smaller than the differential AC Stark shift between the $m_J = -8$ and $m_J = -7$ states. 

After loading a balanced bilayer array, the powers of the two optical standing waves are ramped up, ensuring that the two layers have the same trap frequencies of typically $(\omega_x,\omega_y,\omega_z) = 2\pi \times (0.5,0.5,140)~\rm{kHz}$. The strong axial confinement results in a harmonic oscillator length $a_{\rm{HO}} = \sqrt{\hbar/m\omega_z}$ of $21.1~\mathrm{nm}$, where $\hbar=h/2\pi$ and $m$ is the atomic mass. We load $4.2\times 10^4$ ultracold  $^{162}$Dy atoms into an array of $42$ bilayers, with a temperature of $1.7~\rm{\mu K}$ determined from the cloud size after ballistic expansion (see Methods) \footnote{For technical reasons, the lifetime measurements presented in Fig.~\ref{fig:loss-measurement} are done at trap frequencies of $(\omega_x,\omega_y,\omega_z) = 2\pi \times (0.7,0.7,153)~\rm{kHz}$, leading to an oscillator length of $20.2~\mathrm{nm}$.}. Subsequently, the interlayer distance $s$ is ramped from $\lambda/4$ to different designated values in $0.5~\rm{ms}$ by changing the frequency of the $\sigma_-$ standing wave. The interlayer distance $s$ is calibrated with Kapitza-Dirac diffraction measurements (see Methods). At the end of each experimental sequence, the atoms are released from the bilayer array within $1~\mathrm{\mu s}$ and are imaged after ballistic expansion. With the small axial magnetic field $B_z$ serving as a guiding field, atoms remain in the $m_J = \pm 8$ states and are imaged by a spin-resolved absorption imaging technique (see Methods). This method allows us to measure the population in each of the two layers simultaneously.

We demonstrate the spatial control over the bilayer geometry to be better than 10 nm by scanning the two layers across each other, and measuring the lifetime of the atoms due to loss. The sharp peak in the loss rate as a function of layer separation in Fig.~\ref{fig:loss-measurement} is essentially the convolution between the density profiles of the two layers. Assuming that loss processes occur at short range, we derive a rate equation for the total loss rate of a layer as a function of interlayer distance $s$
\begin{equation}
\Gamma_{\rm{3b}}=\Gamma_{\rm{intra}}+\Gamma_{\rm{inter}}e^{-\frac{1}{3} (\frac{s}{\sigma_z})^2} = \frac{\dot{N}_\mathrm{tot}}{N_\mathrm{tot}}
\label{eq:LossRate}
\end{equation}
where $N_\mathrm{tot}$ is the total number of atoms in a layer, and $\sigma_z$ is the root-mean-square (RMS) thickness of each layer. The loss rate contains both an intra- and an interlayer contribution, and the loss process is assumed to be three-body recombination (see Methods). For spin-independent three-body collisions and thermal clouds, we expect $\Gamma_{\rm{intra}}=\Gamma_{\rm{inter}}$.
Unexpectedly, we observed about a fiftyfold increase in loss rate when the two layers are overlapped, which implies that three-body recombination involving mixed spin states is much faster than recombination of three atoms all in the same spin state. This strongly-enhanced loss feature serves as a highly sensitive monitor for the density overlap between the two layers, while fitting the loss curve determines the thickness of each layer $\sigma_z = 18.6~\rm{nm}$.  This is slightly larger than the RMS size of the harmonic oscillator ground state $a_{\rm{HO}}/\sqrt{2} = 14.3~\rm{nm}$, and can be explained by a small fraction of atoms in excited states of the axial potential \footnote{By taking into account thermal excitation into the higher axial vibrational levels due to finite temperature $T=5.5 ~\mu$K, the thickness of each layer is estimated to be $\sigma_z = \sqrt{\frac{ \sum_{n=0}^\infty (2n+1)e^{-n\hbar\omega_z /k_B T}}{\sum_{n=0}^\infty e^{-n\hbar\omega_z /k_B T}}}\frac{a_{\rm{HO}}}{\sqrt{2}} \approx 1.31  a_{\rm{HO}}/ \sqrt{2} \approx 18.8~\rm{nm}$. Note that the data in Fig.~\ref{fig:loss-measurement} were taken at higher temperature than the other data.}. The observed losses in the two layers are almost equal, implying equal loss rates for three-body collisions involving one spin-up and two spin-down atoms, or vice versa. With this technique of scanning two layers across each other, we have mapped out an atomic density distribution with a resolution much better than 10 nm. This resolution is smaller than previous results of approximately $30~\rm{nm}$ \cite{mcdonald2019superresolution} or $11~\rm{nm}$ \cite{subhankar2019nanoscale} achieved by the dark-state super-resolution microscopy.

We conclude from the loss measurement that for $s\gtrapprox50~\mathrm{nm}$ we can regard the layers as coupled only by long-range dipolar forces. The dipolar energy $U_{\rm{dd}}/h$ between two Dy atoms with opposite spins at this separation is $20~\mathrm{kHz}$. This geometry now allows us to study dipolar physics in new regimes. In the following, we present two out-of-equilibrium experiments which demonstrate strong interlayer dipole-dipole interactions.

\begin{figure}
\includegraphics[width=1.04\columnwidth]{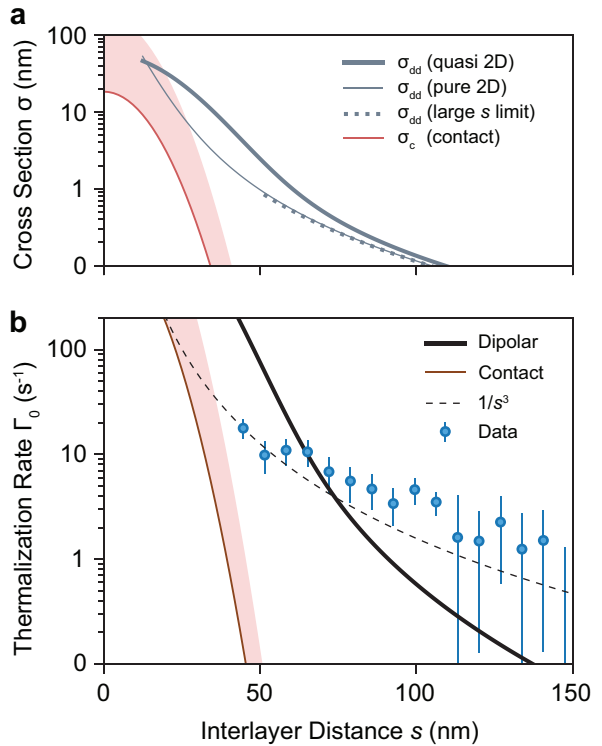}
    \caption{\textbf{Observation of interlayer thermalization.} 
    \textbf{a}, Interlayer elastic scattering cross sections as functions of separation $s$ calculated using the Born approximation. The grey curves correspond to dipolar cross sections for infinitely-thin $\sigma_z = 0$ layers (thin grey) along with its large interlayer distance limit $ks\gg 1$ (dashed, following Eq.~\ref{eq:BornApprox}), and for layers with finite thickness $\sigma_z = 14.9~\rm{nm}$ (quasi 2D, solid grey). The red curve is for simple contact interactions at the background scattering length (red, quasi 2D), and the shaded area corresponds to a 10 times enhanced cross section. 
    \textbf{b}, Observed thermalization rates $\Gamma_0$ obtained from the pseudo-exponential fits. The black and red solid lines show the expected thermalization rate from dipolar and contact interactions (see Methods). The dotted line is for reference and is proportional to $1/s^3$.
    }
\label{fig:interlayer-thermalization}
\end{figure}

One novel experiment is energy-transfer via dipolar interactions, or sympathetic cooling between two atomic systems separated by vacuum \cite{renklioglu2016heat, charalambous2019heat}.  Each layer is heated up by the fluctuating magnetic field created by the dipoles in the other layer.  For equal temperatures, in detailed balance, the heat flows cancel. For unequal temperatures, the dipolar fluctuations cause thermalization.  Fig.~\ref{fig:interlayer-thermalization}(b) shows the experimental results. 

We experimentally create a controlled temperature difference between the two layers by heating up the $\sigma_+$ layer via a parametric drive by modulating the $\sigma_+$ light intensity at twice the transverse trap frequency for $30~\mathrm{ms}$, followed by a $5~\rm{ms}$ hold to ensure any residual breathing motion is damped out. This procedure prepares the $\sigma_+$ layer at  $3.9~\rm{\mu K}$ and leaves the $\sigma_-$ layer at $1.7~\rm{\mu K}$.  We then adjust the interlayer distance over $0.5~\rm{ms}$ and monitor the temperature evolution. We fit the temperature difference between two layers to a pseudo-exponential decay $\frac{\d \Delta T}{\d t} = -\Gamma_0\frac{N(t)}{N(0)}\Delta T$ to obtain the interlayer thermalization rate $\Gamma_0$, where $N(t)$ accounts for the measured particle number decay due to inelastic collisions (see Methods).  Fig.~\ref{fig:interlayer-thermalization}(b) shows that the thermalization rate strongly drops with interlayer distance.

We can estimate the interlayer collision rate as $n_{\rm{2D}} \sigma_{\rm{dd}} v_{\rm{rel}}$, where $n_{\rm{2D}}$ is the 2D density distribution, and $\sigma_{\rm{dd}}$ is the cross section for two dipolar atoms passing each other at the separation $s$.  Using the Born approximation, we calculate the elastic cross section between two atoms in thin layers separated by a distance $s$ (see Methods), and the analytic large-$s$ limit is 
\begin{equation}
    \sigma_{\rm{dd}}^{(\rm{2D})} = a_{\rm{dd}}^2\frac{\pi}{k^2 s^3}
\label{eq:BornApprox}
\end{equation}
Here, $ a_{\rm{dd}} = 10.2$~nm is the dipolar length and $k$ is the relative momentum between the colliding particles.  For $s=75$~nm, the quasi-2D cross section $\sigma_{\rm{dd}}= 0.38$~nm (see Fig.~\ref{fig:interlayer-thermalization}(a)). With a typical 2D peak density of $n_{\rm{2D}} \approx 1.3\times 10^{9}$~cm$^{-2}$ and a thermal velocity of $2.1~\rm{cm/s}$ one obtains an interlayer collision rate of $100~\rm{s}^{-1}$. The observed thermalization times are much slower, around 160 ms (rate of $6~\rm{s}^{-1}$). This can be fully accounted for by the anisotropy of dipolar scattering, which is peaked in the forward direction and reduces the effective cross section by a factor of 6, and by multiple averaging arising from the inhomogeneity of our sample (see Methods). In  Fig.~\ref{fig:interlayer-thermalization}(b), we compare the observed thermalization rates to calculations. They don't have any adjustable parameters and fully take into account the momentum and angular dependence of dipolar scattering and the finite thickness of the layer. The calculations are in semi-quantitative agreement with the observations.  The drop-off of the thermalization rate is much weaker than the steep exponential decrease in density overlap, and therefore in the contact interactions between the two layers.  This is clear evidence for purely dipolar collisions in the range of $50$ to $100~\rm{nm}$ interlayer distances.

The observed dependence on $s$ roughly follows a $1/s^3$ dependence which is less steep than predicted. This is possibly due to the assumptions of the theory based on purely dipolar binary collisions.  For small $s$, there can be an interference term with s-wave contact interactions and a contribution from non-universal short-range dipolar s-wave scattering \cite{ronen2006dipolar} which is not included in the Born approximation.  The largest separations $s$ studied are comparable to the interparticle separation and the binary collision approximation may no longer be accurate, i.e. there are now more then two particles interacting with each other.

\begin{figure}
    \centering
    \includegraphics[width=\columnwidth]{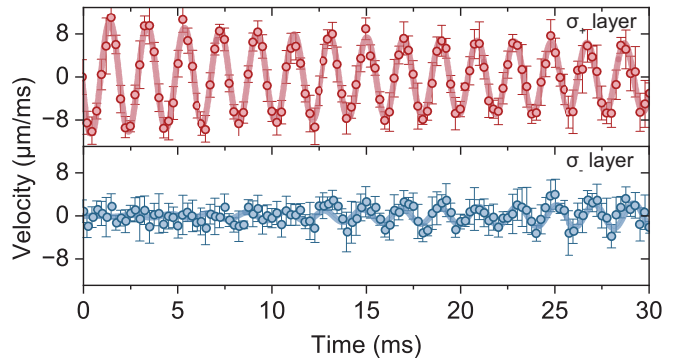}
    \caption{
    \textbf{Observation of coupled oscillations of the two layers at 62 nm interlayer distance.}
   The center-of-mass oscillation of the $\sigma_+$ layer is excited by suddenly switching off a displacement force. The $\sigma_-$ layer oscillates due to dipolar coupling.}
    \label{fig:drag-experiment}
\end{figure}

In the second experiment, we look for coupled collective oscillations of the bilayer system.  Several theoretical papers \cite{Matveeva11, Huang10} predicted the coupling of transverse oscillations by the mean dipolar field between the layers.  Indeed, when we excite transverse oscillations in one layer, we find that they cause oscillations of the other layer (Fig.~\ref{fig:drag-experiment}).  Experimentally,  after loading a balanced bilayer array and adjusting the interlayer distance to a designated value in $0.5~\mathrm{ms}$, we adiabatically displace the $\sigma_+$ layer along the transverse direction $\mathbf{y}$ in $10~\mathrm{ms}$ using an extra laser beam with $\sigma_+$ polarization. This beam, blue-detuned from the $626~\rm{nm}$ transition by $458~\rm{MHz}$, is misaligned from the atoms by about one beam waist, and almost only displaces the atoms in the $\sigma_+$ layer. A sudden switch-off of the displacement beam hence creates a center-of-mass oscillation of the $\sigma_+$ layer at the transverse trap frequency of $500~\mathrm{Hz}$ with an adjustable amplitude ranging from $0$ to $8~\mathrm{\mu m}$ depending on the final power of the beam. As a function of hold time, the in-trap velocity of each layer is obtained from time-of-flight images to reveal how momentum is transferred between layers. The displacement beam also displaces the $\sigma_-$ layer due to crosstalk, but 100 times less. The crosstalk could be observed only at much larger oscillation amplitudes than shown in Fig.~\ref{fig:drag-experiment}.

Fig.~\ref{fig:drag-experiment} shows the time evolution of the velocity of each layer as obtained from ballistic expansion images. The harmonic oscillation of the $\sigma_+$ layer shows damping whereas the $\sigma_-$ layer starts at rest and shows a growing in-phase oscillation.  Our observation is in contrast to the theoretical treatments \cite{Matveeva11, Huang10}, where the mean-field coupling potential would cause a beat note, which is initially an oscillation 90 degrees out of phase.  Furthermore, the predicted mean-field coupling \cite{Matveeva11, Huang10} results in a normal-mode splitting of less than $1~\rm{Hz}$, which is too slow to be observed on the experimental time scale.  Our observation is fully consistent with a friction force caused by dipolar collisions.  First, frictional coupling between harmonic oscillators causes an in-phase oscillation of the driven oscillator.  Second, the time-constant for the damping of the relative motion between the two layers of $25~\rm{ms}$ is similar to the observed interlayer thermalization times.  These observations establish dipolar coupling between two layers which are completely physically separated.

When we explored the coupled oscillations for longer times and for larger amplitudes and separations,  we found that the observations depended critically on a precise matching of the potentials of the two bilayers.  Non-isotropic radial confinement could cause two-dimensional motion of the layers and Lissajous figure type orbits. Nevertheless, all observations showed an initial in-phase oscillation of the $\sigma_-$ layer consistent with a frictional force.

In conclusion, we have presented a new optical super-resolution technique which allows the positioning of atoms at separations of 50 nm and below.  Using higher laser power and tighter confinement, even 10 nm separations should be possible.  We have demonstrated this technique by studying a bilayer of Dy atoms, and have observed strong dipolar interactions via interlayer thermalization and coupled collective oscillations.  The ``dipolar drag'' observed here has common features with Coulomb drag studied in bilayer semiconductors \cite{seamons2009coulomb}.

These results open up many new directions for research. Regarding the bilayer, lower temperatures should lead to strong correlations between the layers beyond a mean-field description.  Adding transverse optical lattices to the layers will create large repulsive interaction energies between pairs on the same lattice site \cite{baranov2012condensed}, but can also realize a system described by attractive interactions between particles and holes analogous to electron-hole pairs in bilayer excitons \cite{eisenstein2004bose}. It is possible to project separate arbitrary potentials into the $\sigma_+$ and $\sigma_-$ layers which could realize twisted bilayer potentials \cite{meng2021atomic} and more general geometries including quasi-crystals. These geometries should allow the study of many phenomena predicted for interacting bilayers \cite{Matveeva11, Huang10, baranov2012condensed, trefzger2009pair, arguelles2007mott, wang2006quantum, koberle2009phonon, macia2014single, hufnagl2013stability, safavi2013quantum}. We are currently applying the super-resolution technique to optical tweezers to place two atoms at 50 nm scale separation.  This will allow the study of superradiance and radiative shifts at separations much smaller than the optical wavelength, and the study of magnetic interactions and spin exchange between two isolated atoms, as done recently with polar molecules \cite{holland2022demand, bao2022dipolar, christakis2022probing, yan2013observation}. The tweezer setup can be generalized to a linear array of atoms alternating in spin-up and spin-down states.  Moving the spin-up atoms back and forth would provide full connectivity along the chain, and realize a spin chain with strong magnetic coupling between nearest neighbors. These ideas can be generalized to higher dimensions.

\textbf{Acknowledgments}  We thank Alan Jamison, Jin Yang, Jiahao Lyu, and Tom De Coninck for 
experimental assistance and discussions, and Jinggang Xiang and Alan Jamison for comments on the manuscript. We acknowledge support from the NSF through the Center for Ultracold Atoms and through Grant No. 1506369, the Vannevar-Bush Faculty Fellowship, and an ARO DURIP grant.

\textbf{Author contributions} L.D., P.B., M.C. designed and constructed the experimental setup, L.D., P.B., M.C., J.d.H., Y.-K.L. carried out the experimental work. All authors contributed to the development of models, data analysis and writing of the manuscript.

\textbf{Competing interests} The authors declare no competing interests.

\bibliography{bilayer}

\clearpage

\setcounter{equation}{0}
\setcounter{figure}{0}
\setcounter{table}{0}
\makeatletter
\renewcommand{\theequation}{S\arabic{equation}}
\renewcommand{\thefigure}{S\arabic{figure}}

\section*{Methods}

\renewcommand{\paragraph}[1]{\textbf{#1}.---}

\paragraph{Sample preparation}
Typically, a Bose--Einstein condensate (BEC) with $5\times 10^4$ $^{162}\rm{Dy}$ atoms in the lowest Zeeman state $m_J = -8$ is prepared by first loading a $1064~\mathrm{nm}$ crossed optical dipole trap (XODT) from the $626~\rm{nm}$ spin-polarized magneto-optical trap \cite{Lunden20}. Evaporative cooling is then performed in a magnetic field of $0.42~\rm{G}$ by ramping down the depth of the XODT. The trap frequencies of the XODT at the end of evaporation are $(\Omega_x, \Omega_y, \Omega_z) = 2\pi\times (149, 43, 136)~\rm{Hz}$. The condensate fraction is about $45\%$ and the temperature is $60~\rm{nK}$.

By following the sequence depicted in Fig.~\ref{fig:potential-cartoon}(d), approximately $4.2\times 10^4$ atoms are subsequently loaded into the bilayer potential. Due to the partial condensation of the cloud and technical heating during the loading, we simply assume a Gaussian distribution in the transverse direction such that the atomic density is
\begin{equation}
    n_{i,\rm{3D}}^{(\pm)}(\rho,z)=
    n_{0,\rm{3D}}
    e^{ -\frac{1}{2}\left(i\frac{\lambda}{2 \sigma_{\rm{ODT}}}\right)^2} 
    e^{-\frac{1}{2}\left(\frac{\rho}{\sigma_\perp}\right)^2}
    e^{-\frac{1}{2}\left(\frac{z\pm s/2}{\sigma_z}\right)^2}
    \label{eq:layerDensity}
\end{equation}
where $i=0, \pm 1, \pm 2, \ldots$ is the index of the bilayer in the whole array, and the superscript $(\pm)$ denotes the $\sigma_+$ or $\sigma_-$ layer. The transverse width (RMS value of x and y) is given by $\sigma_\perp = \sqrt{\frac{k_B T}{m\omega_\rho^2}}=3.0~\rm{\mu m}$ with $\omega_\rho = 2\pi\times 500$~Hz and $T = 1.7~\mu$K. The RMS thickness $\sigma_z$ relates to the oscillator length as $\sigma_z = a_{\rm{HO}}/\sqrt{2}$, therefore in the tight direction $\sigma_z = \sqrt{\frac{\hbar}{2m\omega_z}} = 14.9~\rm{nm}$ for $\omega_z = 2\pi\times 140~\rm{kHz}$.

We measured $\sigma_{\rm{ODT}}$ by a matter-wave focusing technique \cite{PhysRevLett.89.270404, PhysRevLett.120.060402} which maps the spatial distribution into momentum distribution which is imaged after ballistic expansion. The procedure is the following: first, the XODT is turned back on and the bilayer potential is switched off suddenly after loading. After a quarter period of oscillation in the XODT, the initial spatial distribution is converted into a momentum distribution. Finally, atoms are released from the XODT for absorption imaging after ballistic expansion for time $t$. The width of the density distribution in the bilayer potential is related to the width of the cloud in the absorption image by $\sigma_{\rm{ODT}}=\sigma_{\text{TOF}}/(\Omega_z t)=4.4~\rm{\mu m}$. This value agrees with a value obtained by in-situ imaging of the bilayer using detuned imaging light. As a result, we populate about $4\sqrt{\pi}\sigma_{\rm{ODT}}/\lambda \approx 42$ bilayers with 700 atoms in each layer of the central bilayer.

The peak density $n_{0,\rm{3D}}$ is expressed in terms of experimentally measured quantities as $n_{0,\rm{3D}}= 3N_{\rm{tot}}^{(\pm)}/V_{\rm{eff}} $ with the effective trap volume of the whole bilayer array $V_{\rm{eff}}=12 \pi^2 \sigma^2_\perp \sigma_z (2\sigma_{\rm{ODT}}/\lambda)$. We assume balanced loading of the bilayers, so the peak 2D density $n_0 = \sqrt{2 \pi}\sigma_z n_{0,\rm{3D}} = 1.3\times 10^{9}$~cm$^{-2}$ is the same in both layers initially.

The bilayers are thermal since the temperature increases to $T=1.7~\rm{\mu K}$ at the end of the loading procedure, which is above the critical temperature for 2D Bose-Einstein condensation of $0.5~\rm{\mu K}$. The energy scale set by the temperature of each layer is much larger than the transverse vibrational energy spacing $ k_B T / \hbar \omega_x = k_B T / \hbar \omega_y \approx 70 \gg 1$, but is small compared to the axial vibrational energy spacing $k_B T / \hbar \omega_z \approx 0.25 <1$. The  axial thermal excitation is on the order of $e^{-\hbar \omega_z/k_B T} = 0.02$ (and $0.18$ for the heated layer at $3.9~\rm{\mu K}$ in the interlayer thermalization experiment).

\paragraph{Control and characterization of interlayer distances}
To control the distance between two layers, we shift the frequency of the $\sigma_- $ optical standing wave, leading to a variable layer displacement due to the accumulated phase shift of the standing wave at the position of the atoms that is distance $L$ away from the retro mirror. With $\delta f= 80~\rm{MHz}$ tuning range of the $\sigma_- $ optical frequency and $1.9~\rm{m}$ of retro-path length, we are able to shift the $\sigma_- $ layers by a distance $ L \cdot \delta f/ f = 375 ~\rm{nm}$ with respect to the $\sigma_+ $ layers, where $f$ is the frequency of the $741~\rm{nm}$ laser. 

\begin{figure*}[t]
\includegraphics[width=0.6\textwidth]{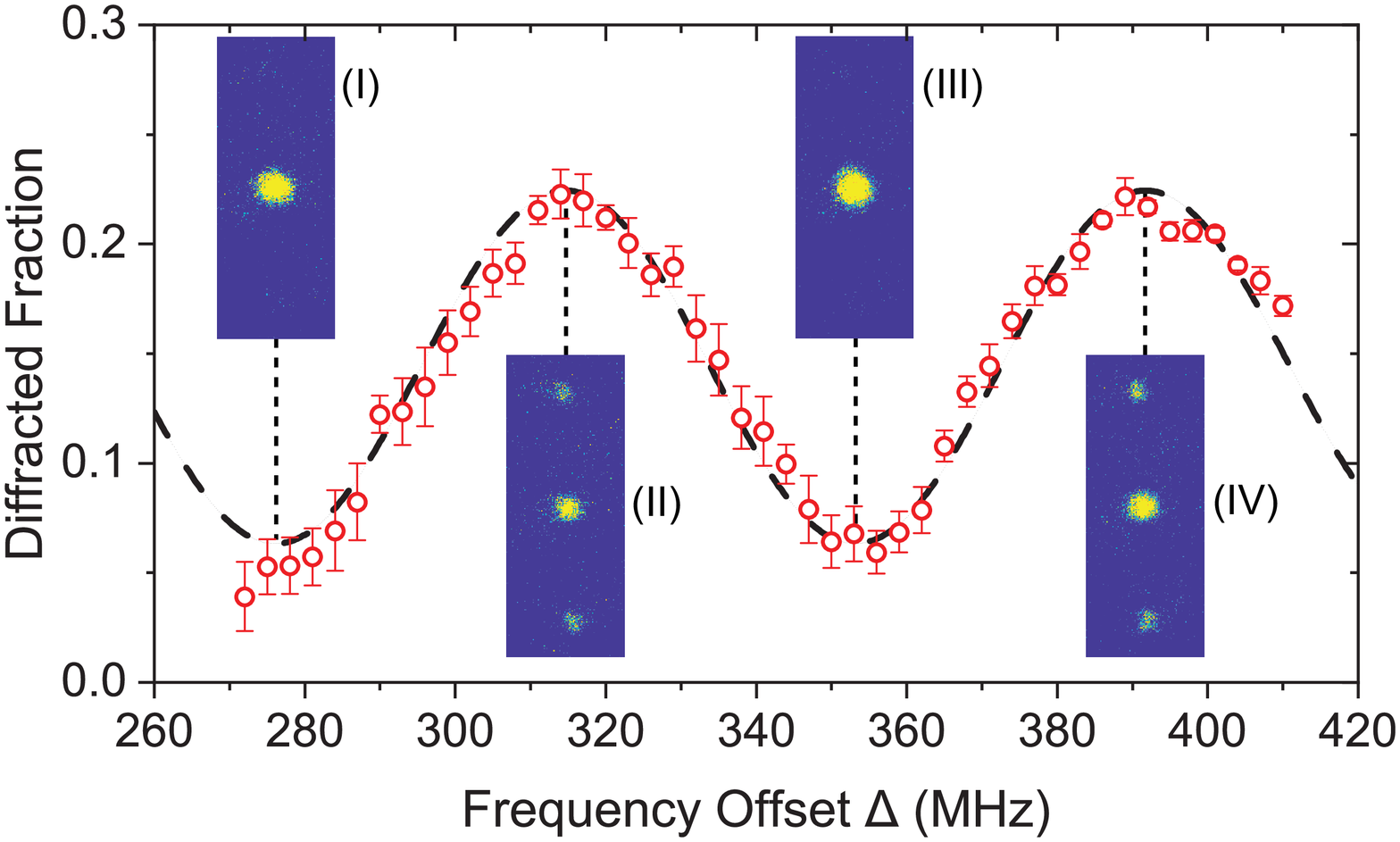}\\
\caption{\textbf{Characterization of the interlayer distance using Kapitza-Dirac diffraction.} 
The 1st-order fraction of the Kapitza-Dirac signals $N_{+1}/N_{\rm{tot}}$ versus relative detuning $\Delta$ with $m_J=-8$ Bose-Einstein condensates polarized along the $\mathbf{x}$ direction.
The Kaptiza-Dirac signal vanishes at interlaced bilayer configurations (I) and (III), whereas it is maximized for overlapped bilayer configurations (II) and (IV). 
}
\label{fig:ApparatusKD}
\end{figure*}

The distance is calibrated using a Kapitza-Dirac experiment \cite{Gould86} in which the diffraction patterns of atoms are used to reveal the structure of the pulsed optical standing waves. This is done  with a BEC in the $m_J = -8$ Zeeman state polarized along the transverse direction $\mathbf{x}$ in a magnetic field of $1.3~\rm{G}$. After the preparation of the condensate, we simultaneously pulse on the two optical standing waves with the same intensity for $\tau=5~\mathrm{\mu s}$, such that most of the population is in the the first order of the Kapitza-Dirac diffraction pattern.  Since the transversely-polarized atomic spins see both circular polarizations as a superposition of ($\sigma_-, \pi, \sigma_+$) light with weights of (1, 2, 1), taking $\mathbf{x}$ as the quantization axis, the light-atom interaction Hamiltonian for the $m_J = -8$ state can be expressed as the superposition of two phase-shifted sinusoidal potentials of the same amplitudes $V_0(x,y) [\sin^2(kz)+ \sin^2(kz+\phi)]$. Fig.~\ref{fig:ApparatusKD} shows the typical first-order Kapitza-Dirac signals in the short pulse limit $V_0(x,y)\tau/h \ll 1$. The result presents an oscillatory behavior as we vary the relative detuning $\Delta$. When the bilayers are in an interlaced configuration ($\phi = \pi /2$), the first-order Kapitza-Dirac signal vanishes. When the bilayers are in an overlapped configuration ($\phi = 0$) the amplitude of the sinusoidal potential is maximized, corresponding to the strongest Kapitza-Dirac signal. 
The resulting oscillation period of the Kapitza-Dirac signal indicates the tuning sensitivity of the interlayer distance to be $4.7~\rm{nm/MHz}$ with respect to the relative laser detuning $\Delta$.

\begin{figure*}
\includegraphics[width=0.9\textwidth]{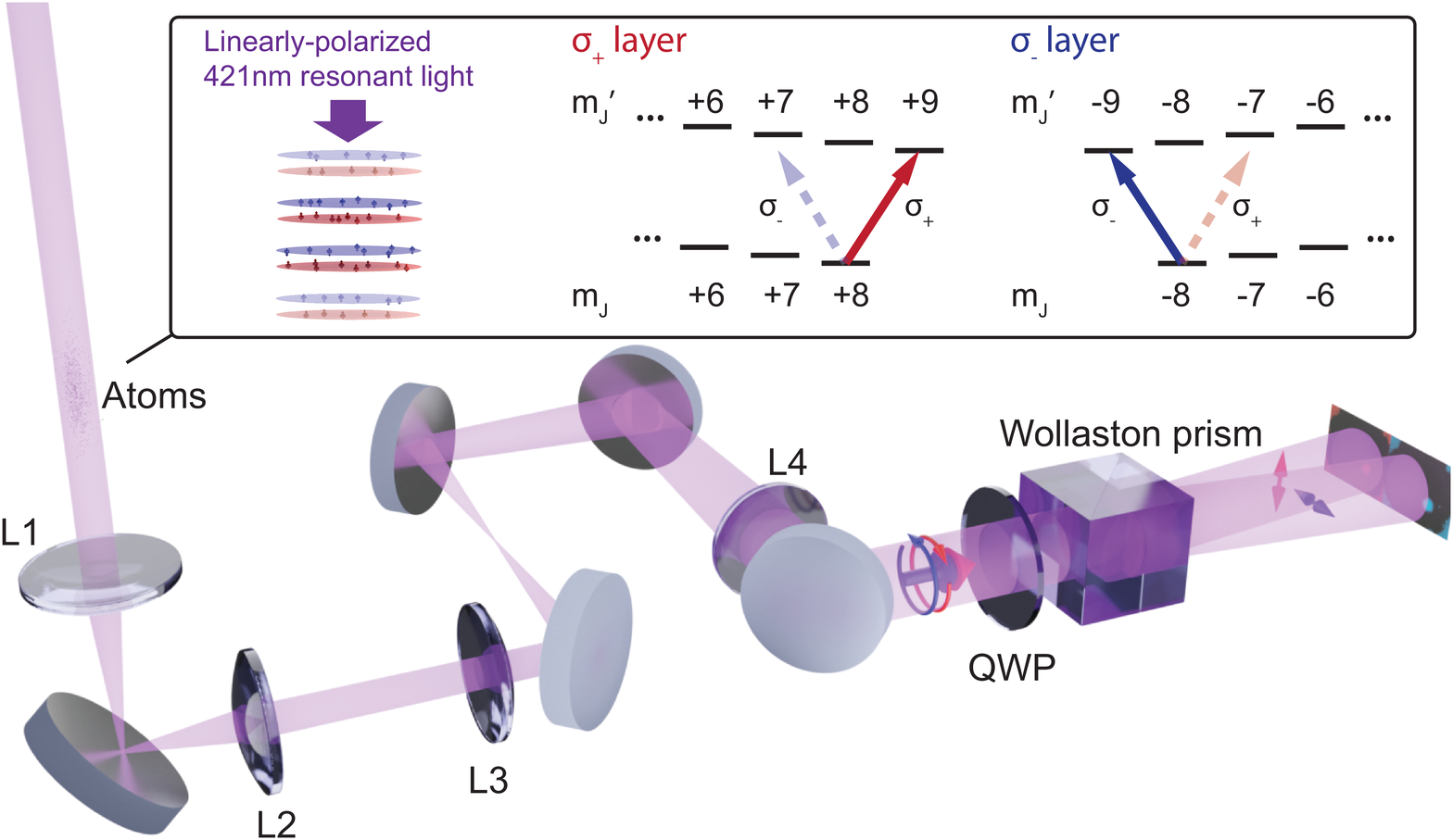}\\
   \caption{\textbf{The spin-resolved absorption imaging scheme.}
   The atoms are exposed to linearly polarized light that is co-propagating with the bilayer optical beams and that is resonant with the $421~\mathrm{nm}$ transition. After the relay lenses ($L1$ and $L2$) and the magnification lenses ($L3$ and $L4$), the $\sigma_+$ and  $\sigma_-$ components of the imaging light are separated using a quarter-wave plate (QWP) and a Wollaston prism. The spatially separated images of the $\sigma_+$ and the $\sigma_-$ layers are recorded by a CMOS camera. 
}
\label{fig:spin-resolved-imaging}
\end{figure*}

\paragraph{Spin-resolved imaging}
Our spin-resolved absorption imaging system operates in the weak saturation limit of the $421~\mathrm{nm}$ cyling transition. It utilizes the big contrast of photon scattering rates of atoms in the stretched $m_J = \pm 8$ Zeeman states for two opposite circular polarizations of light. The resonant imaging light addressing the $421~\mathrm{nm}$ transition is linearly polarized along $\mathbf{x}$, and propagates along the axial direction $\mathbf{z}$. Taking $\mathbf{z}$ as the quantization axis, the imaging light contains equal amount of $\sigma_-$ and $\sigma_+$ polarization components. Due to the big difference between the Clebsch--Gordan coefficients for the $\ket{J=8, m_J=-8}\rightarrow\ket{J'=9, m'_J=-9}$ and the $\ket{J=8, m_J=-8}\rightarrow\ket{J'=9, m'_J=-7}$ electric dipole transitions, the $\sigma_+$ photons are predominantly scattered by the atoms in the $\sigma_+$ layer, whereas the $\sigma_-$ photons are predominantly scattered by the atoms in the $\sigma_-$ layer. The two polarization components are then spatially separated by a $1^{\circ}$ angle via a quarter-wave plate and a Wollaston prism (see Fig.~\ref{fig:spin-resolved-imaging}), leading to two nearly-independent imaging channels for the $\sigma_+$ and the $\sigma_-$ layers on the camera. The duration of the imaging pulse is adjusted to reduce  optical pumping which would lead to crosstalk between the two imaging channels.

\paragraph{Intra- and interlayer loss rates}
Here we provide the details of the model for determining the three-body loss rates for separated and overlapping layers. For technical reasons, these experiments were carried out with different parameters.  Instead of loading from a BEC, the bilayers here were loaded from an ultracold thermal cloud at $T = 172~\rm{nK}$ with no condensate fraction. The thermal cloud follows a Gaussian density distribution with widths of $(\tilde{\sigma}_x, \tilde{\sigma}_y, \sigma_{\rm{ODT}}) = \sqrt{\frac{k_B T}{m}}(\frac{1}{\Omega_x}, \frac{1}{\Omega_y}, \frac{1}{\Omega_z}) = (3.1, 11.0, 3.5)~\rm{\mu m}$ \cite{ketterle1999making}, and thus approximately $33$ copies of the bilayer are created. The density distribution in each layer follows Eq.~\ref{eq:layerDensity}. The trap frequencies of the bilayer were $(\omega_x ,\omega_y, \omega_z )=2\pi\times(0.7,0.7,153)~\rm{kHz}$, corresponding to an axial oscillator lengths of $a_{\rm{HO}} = \sqrt{\hbar/m\omega_z} = 20.2~\rm{nm}$. The typical post-loading peak density is $n_{0,\rm{3D}}=3.8\times 10^{14} ~\rm{cm^{-3}}$.

For separated layers, the local density $n$ decays by three-body loss according to $\d n/\d t = -\beta_{\rm{intra}} n^3$.  By integrating over the cloud and layers we obtain the first term of the rate equation
\begin{equation}
    \frac{dN_{\rm{tot}}}{dt}=
    -\beta_{\rm{intra}}\frac{N_{\rm{tot}}^3}{V_{\rm{eff}}^2}
     - \beta_{\rm{inter}}\frac{N_{\rm{tot}}^3}{V_{\rm{eff}}^2}e^{-\frac{1}{3}(\frac{s}{\sigma_z})^2}
\label{eq:rateEquation}
\end{equation}
with $V_{\rm{eff}}=12 \pi^2 \sigma^2_\perp \sigma_z (2\sigma_{\rm{ODT}}/\lambda)$. The second term characterizes interlayer loss when the layers overlap with a rate constant $\beta_{\rm{inter}}$. The prefactor 1/3 in the exponent assumes three-body loss. For two-body loss, it would be 1/4. Although two-body spin relaxation between $m_J = \pm 8$ states becomes energetically possible when the layers partially overlap, it requires a higher-order process and should be negligible. The three-body recombination rate is proportional to the square of the density. Assuming that three-body collisions are independent of the spin state on a microscopic level, one would naively expect a fourfold increase of $\Gamma_{\rm{3b}}$ when the layers fully overlap --- this is correct only in the case of a two-component Bose condensate.   
 For spin-independent three-body collisions and a Bose condensate, we define $\beta =\beta_{\rm{intra}}$.  We then get  $\beta_{\rm{inter}} = 3 \beta$. For thermal clouds we have $\beta_{\rm{intra}}  = g^{(3)}(0) \beta$ and $\beta_{\rm{inter}} = 3 g^{(2)}(0) \beta = \beta_{\rm{intra}}$. and therefore an expected twofold increase of the loss rate for fully overlapping layers.

Fitting the initial loss rates $\dot{N}_{\rm{tot}}/N_{\rm{tot}}$ as a function of the interlayer distances $s$ using Eq.~(\ref{eq:LossRate}) gives the intralayer three-body loss coeffcient $\beta_{\rm{intra}} = 9.0\times 10^{-28} ~\rm{cm^6 /s} $, the interlayer three-body loss coefficient $\beta_{\rm{inter}} = 4.8\times 10^{-26} ~\rm{cm^6 /s}$, and the RMS thickness of each layer $\sigma_z = 18.6~\mathrm{nm} \approx 1.3 a_{\rm{HO}}/\sqrt{2}$. The intralayer three-body loss coefficient is of the same order of magnitude compared to the results in previous works \cite{bottcher2019dilute, FamaThesis2018} measured around $5~\rm{G}$ magnetic fields away from Feshbach resonances.

\paragraph{Excitation and measurement of center-of-mass oscillations}
Oscillations of the $\sigma_+$ layer are excited using a circularly-polarized beam blue-detuned by $458~\rm{MHz}$ from the $626~\rm{nm}$ transition. The focus of the beam is misaligned along the $\mathbf{y}$ direction with respect to the atoms, causing a force that displaces the atoms along $\mathbf{y}$ due to the AC Stark shift gradient. The spin-selectivity of the beam due to its circular polarization guarantees that it predominantly addresses the $\sigma_+$ layer. By adiabatically ramping up the beam in $10~\rm{ms}$, we displace the $\sigma_+ $ layer for various distances controlled by the final power of the beam. A center-of-mass oscillation is excited by suddenly switching  off the beam. We determined that the oscillation amplitude of the $\sigma_-$ layer caused by the beam is $1.2(5)\%$ compared to that of the $\sigma_+$ layer by measuring the amplitudes with only the $\sigma_+$ or $\sigma_-$ layer loaded. This is consistent with the Clebsch-Gordon coefficients for the two polarizations.

\paragraph{Calculation of $\sigma_{\rm{dd}}$}
The theoretical curves of the thermalization rate $\Gamma_0$ in Fig~\ref{fig:interlayer-thermalization} use the distance-dependant interlayer dipolar cross section $\sigma_{\rm{dd}}$ of two atoms confined in two different layers. To compute this quantity, we use the Born approximation similar to Ref.~\cite{Pasquiou10, dypole2023dipolar}. In the center-of-mass frame of two particles (labeled ${1,2}$) confined in two layers separated by distance $s$, the axial potential is reduced to a single harmonic oscillator potential $\hat V_{\rm{HO}} = \frac{1}{2}\mu\omega_z^2(z-s)^2$, described by the relative axial coordinate $z = z_2-z_1$ and a reduced mass $\mu = m/2$. Writing the transverse part of the wavefunction in the form of
\begin{equation}
    \Psi(\bm{\rho}) = e^{ik\mathbf{u_i}\cdot\bm{\rho}} + e^{i\pi/4}f(k, \theta)\frac{e^{ik \rho}}{\sqrt{\rho}}
\end{equation}
leads to a scattering amplitude
\begin{equation}
\label{eq:scattering_amplitude}
f(k, \theta) = \frac{\mu}{\hbar^2}\frac{-1}{2\sqrt{2}\pi^{3/2}}\frac{1}{\sqrt{k}}\int\d q_z\mathcal{H}(-q_z)\mathcal{V}(\mathbf{q})
\end{equation}
with $\mathbf{u_i}$ being the direction of the incident plane wave, $\mathbf{u_\rho}$ being the direction of the scattered wave, $\cos\theta =\mathbf{u_i}\cdot\mathbf{u_\rho}$ being the scattering angle, and $\mathbf{q} = k(\mathbf{u_\rho} - \mathbf{u_i}) + q_z\mathbf{u_z}$. Here $\mathcal{H}(q_z) = e^{-iq_zs-\sigma_z^2q_z^2}$ is the Fourier transform of the harmonic oscillator ground state density of the two-particle potential $\hat V_{\rm{HO}}$, and 
$\mathcal{V}(\mathbf{q}) = 4\pi\frac{\hbar^2}{\mu}a_{\rm{dd}}(\frac{q_z^2}{\left|\mathbf{q}\right|^2} - 1)$ is the Fourier transform of the dipole-dipole interaction. Integrating over the angle $\theta$ leads to the 2D interlayer dipolar cross section
\begin{equation}
    \sigma_{\text{dd}} = \int_{0}^{2\pi}\d\theta\left|f(k,\theta)\right|^2
    \label{eq:sigmaQuasi2D}
\end{equation}
This is the quasi-2D result presented in Fig.~\ref{fig:interlayer-thermalization}(a).

An analytic form of the cross section can be obtained in the pure 2D limit $\sigma_z = 0$ where the thickness of each layer is regarded as negligible. In this limit, the integral involved in the scattering amplitude can be simplified as $\int\d q_z e^{iq_z s}\left(\frac{q_z^2}{q_\rho^2+q_z^2}-1\right) = -q_\rho \pi e^{-q_\rho s}$. Since $q_\rho = k\left|\mathbf{u_\rho}-\mathbf{u_i}\right|$, the integral for the 2D cross section is reduced to $\int_{0}^{2\pi}\d\theta q_\rho^2e^{-2q_\rho s} = 2k^2 \int_{0}^{2\pi}\d\theta (1-\cos\theta)e^{-2\sqrt{2}ks\sqrt{1-\cos\theta}}\simeq 2k^2\int_0^{\infty}\d\theta\theta^2e^{-2ks\theta}$. For large interlayer distance $ks\gg 1$, the pure 2D dipolar cross section asymptotes to
\begin{equation}
    \sigma_{\rm{dd}}^{\text{(2D)}} = a_{\rm{dd}}^2 \frac{\pi}{k^2 s^3}
    \label{eq:sigmaPure2D}
\end{equation}
This is the analytic result in Eq.~\ref{eq:BornApprox} and the dashed curves in Fig.~\ref{fig:interlayer-thermalization}(a).

It is known that the Born approximation breaks down for overlapping layers $s=0$ \cite{ticknor2009two} due to the $1/r^3$ divergence of the dipole-dipole interaction. In order for the Born approximation to be valid, one requires the scattered part of the wavefunction to be small. A sufficient condition for this is
\begin{equation}
    \frac{\mu}{\sqrt{2\pi}\hbar^2}\left|\int \d^2\rho_1\frac{V_{\rm{dd}}(\bm{\rho}_1)}{\sqrt{k\left|\bm{\rho}_1\right|}}\right|\ll 1
\end{equation}
This condition is fulfilled when $ks^3\gg 1.13 a_{\rm{dd}}^2$. For our temperature range, this is satisfied when $s\gg 13$~nm. 

\begin{figure*}
    \centering
    \includegraphics[width=\columnwidth]{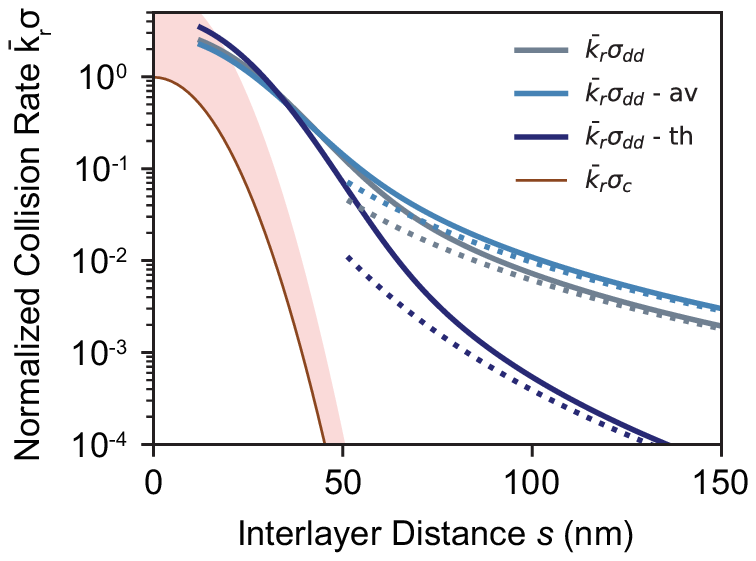}
    \caption{
    \textbf{Calculation of normalized collision rates.} 
    Dimensionless, density normalized collision rates $\bar k_r\sigma$ computed for contact and dipolar potential. $\bar k_r$ is the mean average momentum $\kappa\sqrt{\pi}/2$. The brown curve correspond to the contact interaction from Eq.~\ref{eq:sigmaQuasi2Dcontact} with the shaded red area indicating a 10 times larger cross section. The dashed gray, light blue and navy curves correspond to dipolar interaction in the pure 2D large-$s$ approximation, Eq.~\ref{eq:sigmaPure2D}, \ref{eq:sigmaPure2Ddipolar_av} and \ref{eq:sigmaPure2Ddipolar_th}, respectively. The solid curves are for the quasi-2D cases, Eq.~\ref{eq:sigmaQuasi2D}, \ref{eq:sigma_av} and \ref{eq:sigma_th}, respectively.
    }
    \label{fig:theoryCrossSection}
\end{figure*}

\paragraph{Calculation of $\sigma_c$}
Eq.~\ref{eq:scattering_amplitude} can also be used to compute the effective quasi-2D cross-section arising from the contact $s$-wave interaction. For a scattering length $a_s$, the effective potential $\mathcal{V}_c(\bm{q}) = \frac{2\pi\hbar^2}{\mu}a_s$ gives
\begin{equation}
    \sigma_c(k) = \frac{4\pi a_c^2}{k\sigma_z^2}e^{-2s^2/\sigma_z^2}
    \label{eq:sigmaQuasi2Dcontact}
\end{equation}
We use the low-field background value of $a_c = 5.9$ nm \cite{tang2015s} in our calculations. The scattering length for collisions between -8 and +8 atoms is not known. The large three-body losses (which asymptotically scales with the forth power of the scattering length \cite{esry1999recombination}) is an indication that the cross section for collisions between -8 and +8 atoms is 5 - 10 times larger. The shaded area in Figs.~\ref{fig:interlayer-thermalization} and \ref{fig:theoryCrossSection} indicates a range of a factor of 10.

\paragraph{Calculation of thermalization rate}
The thermalization rate $\Gamma$ typically derives from the collision rate $\gamma$. The latter is given by the product of density $n$, the cross section $\sigma$ and the average velocity $\bar{v}_r$. The thermalization rate is then usually obtained by dividing this quantity by the number of collisions necessary to reach thermalization. However in our case, although the dipolar potential in the $x-y$ plane is isotropic, the cross section $\sigma_{\text{dd}}(k)$ is momentum dependent and highly anisotropic since the scattering amplitude $f(k,\theta)$ depends strongly on each of its variables. It is then necessary to take the full momentum  distribution into account to compute the collision rate as in Ref.~\cite{anderlini2005model}, which leads to the definition of an effective averaged cross section $\sigma_{\text{av}}$ for collisions
\begin{equation}
\gamma = \frac{1}{2^{3/2}} n_{0,\rm{eff}} \bar v_r\sigma_\text{av}
\label{eq:collision_rate}
\end{equation}
where $n_{0,\rm{eff}} = n_0^{(-)}T^{(-)}/\bar T$ is the effective 2D peak density, and each three $\sqrt{2}$ factors comes from the averaging of the rate along the  two radial directions of the pancake and the discrete average over the stack of bilayer given the initial width of the loaded  cloud. Introducing $\kappa = \sqrt{\frac{4k_B\bar T m}{\hbar^2}} = (2/\sqrt{\pi})(m\bar{v}_r/\hbar)$ we have
\begin{equation}
\sigma_{\text{av}} = \frac{4}{\sqrt{\pi}}\frac{1}{\kappa^3}\int_0^\infty dk\sigma(k)k^2e^{-k^2/\kappa^2}
\label{eq:sigma_av}
\end{equation}
The thermalization rate $\Gamma$ also incorporates the anisotropic redistribution of momentum between the two clouds. Forward scattering is less efficient for thermalization than scattering at larger angles.  Therefore, we introduce the effective thermalization cross section $\sigma_\text{th}$ such that
\begin{equation}
    \Gamma = \frac{1}{2^{3/2}}n_{0,\rm{eff}}\bar v_r\sigma_\text{th}
\label{eq:thermalization_rate}
\end{equation}
with
\begin{equation}
    \sigma_{\text{th}} = \frac{4}{\sqrt{\pi}}\frac{1}{\kappa^5}\int_0^\infty dk  k^4e^{-k^2/\kappa^2}  \int_0^{2\pi} d\theta (1-\cos \theta) \left|f(k,\theta)\right|^2 
    \label{eq:sigma_th}
\end{equation}
The definition is such that $\sigma_{\text{av}} = \sigma$ for a momentum independent scattering amplitude. The ratio $\sigma_{\text{av}}/\sigma_{\text{th}}$ is the number of collision for thermalization. This ratio is 2/3 for an isotropic and momentum-independent scattering in 2D. For dipolar scattering, the ratio is much larger and depends on momentum and interlayer separation.

The contact interaction in 2D is a particular case where the ratio is 1, and the averaged cross section is also equal to the non-averaged one taken at the mean relative momentum $\bar k_r = \kappa\sqrt{\pi}/2$, namely $\sigma_{c, \text{av}}(\kappa) = \sigma_{c, \text{th}}(\kappa) = \sigma_c(\bar k_r)$. The momentum-independent quantity $\bar k_r\sigma_c$ is plotted in Fig.~\ref{fig:theoryCrossSection}. The impact of anisotopy is striking for the dipolar case. In the pure 2D and large distance $\kappa s\gg 1$ limits, the effective cross sections are
\begin{equation}
    \sigma_{\text{dd, av}} = a_{\rm{dd}}^2\frac{2\pi}{\kappa^2 s^3}
    \label{eq:sigmaPure2Ddipolar_av}
\end{equation}
and
\begin{equation}
    \sigma_{\text{dd, th}} = a_{\rm{dd}}^2\frac{3\pi}{\kappa^4 s^5}
    \label{eq:sigmaPure2Ddipolar_th}
\end{equation}
which are also plotted in Fig.~\ref{fig:theoryCrossSection}.  The cross section $\sigma_{\text{dd, th}}\propto 1/s^5$ falls off with distance more rapdily than the dipolar potential $\propto  1/s^3$ since the larger the distance, the more pronounced forward scattering becomes. The red and black curves shown in Fig.~\ref{fig:interlayer-thermalization}(b) are $\Gamma_c$ and $\Gamma_{\text{dd}}$, respectively.

\paragraph{Thermalization fit}
The thermalization rate $\Gamma_{\rm{dd}}$ depends on $n_{0}^{(-)}T^{(-)}/\bar T$, $\sigma_{\rm{dd, th}} \propto 1/k_r^4$ and $v_r\propto k_r$. The last two quantities purely depend on the relative momentum $k_r\propto  \sqrt{\bar T}$ which is constant during the thermalization. The density scales with temperature as $n_{0}^{(-)}\propto N_{\rm{tot}}^{(-)}/T^{(-)}$, hence $n_{0}^{(-)}T^{(-)}/\bar T \propto N_{\rm{tot}}^{(-)}/\bar T$ which leaves   $\Gamma_{\rm{dd}}$ independent of the temperature difference between the two layers $\Delta T = T^{(+)} - T^{(-)}$. Therefore the only changing variable left out is the total atom number, and the differential equation for $\Delta T$ is
\begin{equation}
    \frac{\d \Delta T}{\d t} = -\Gamma(t)\Delta T = -\Gamma_0\frac{N_{\rm{tot}}^{(\pm)}(t)}{N_{\rm{tot}}^{(\pm)}(0)}\Delta T
\end{equation}
We observe losses attributed to three-body recombination with a timescale of $100~\rm{ms}$, and therefore we take them into account for fitting $\Delta T$. Assuming a rate equation $\frac{\d N}{\d t} = -\Gamma_n N^3$ leads to the evolution of the temperature difference
\begin{equation}
    \Delta T(t) = \Delta T_0 e^{\frac{\Gamma_0}{\Gamma_n N_0^2}\left(1-\sqrt{1+2\Gamma_n N_0^2 t}\right)}
\end{equation}
from which we can extract the interlayer thermalization rate $\Gamma_0$, plotted in Fig.~\ref{fig:interlayer-thermalization}.
\\\\\\\\\\\\\\\\
\\\\\\\\\\\\\\\\
\\\\\\\\\\\\

\newpage

\end{document}